\documentclass[%
 aip,
 jmp,%
 amsmath,amssymb,
 reprint,%
]{revtex4-1}

\usepackage{graphicx}
\usepackage{dcolumn}
\usepackage[utf8]{inputenc}
\usepackage[none]{hyphenat}
\usepackage{bm}
\usepackage{amsmath}
\usepackage{subfigure}
\usepackage[x11names,table,xcdraw]{xcolor}

\begin{document}


\title[ ]{Impact of growth conditions on the domain nucleation and domain wall propagation in Pt/Co/Pt stacks} 



\author{Cynthia P. Quinteros}
\altaffiliation{Present address: Zernike Institute for Advanced Materials, University of Groningen, 9747 AG Groningen, The Netherlands}
\email{cpquinterosdominguez@gmail.com}
\affiliation{Instituto de Nanociencia y Nanotecnología CNEA-CONICET, Centro Atómico Bariloche, Av. E. Bustillo 9500, (R8402AGP) S. C. de Bariloche,
Río Negro, Argentina} 

\author{María José Cortés Burgos}
\affiliation{Instituto de Nanociencia y Nanotecnología CNEA-CONICET, Centro Atómico Bariloche, Av. E. Bustillo 9500, (R8402AGP) S. C. de Bariloche,
Río Negro, Argentina} 
\affiliation{Instituto Balseiro, Universidad Nacional de Cuyo - CNEA, Av. Bustillo 9500, R8402AGP, S. C. de Bariloche, Rio Negro, Argentina}

\author{Lucas J. Albornoz}
\affiliation{Instituto de Nanociencia y Nanotecnología CNEA-CONICET, Centro Atómico Bariloche, Av. E. Bustillo 9500, (R8402AGP) S. C. de Bariloche,
Río Negro, Argentina}
\affiliation{Instituto Balseiro, Universidad Nacional de Cuyo - CNEA, Av. Bustillo 9500, R8402AGP, S. C. de Bariloche, Rio Negro, Argentina}
\affiliation{Laboratoire de Physique des Solides, Université Paris-Sud, Université Paris-Saclay, CNRS, UMR8502, 91405 Orsay, France}

\author{Javier E. Gómez}
\affiliation{Instituto de Nanociencia y Nanotecnología CNEA-CONICET, Centro Atómico Bariloche, Av. E. Bustillo 9500, (R8402AGP) S. C. de Bariloche,
Río Negro, Argentina}

\author{Pablo Granell}
\affiliation{Centro de Micro- y Nanoelectrónica Del Bicentenario, Instituto Nacional de Tecnología Industrial, Av. Gral Paz 5445, San Martín Buenos Aires, B1650KNA, Argentina}

\author{Federico Golmar}
\affiliation{ECyT, Universidad de San Martín, Av. 25 de Mayo y Francia (1650) San Martín, Argentina}

\author{María Luján Ibarra}
\affiliation{ECyT, Universidad de San Martín, Av. 25 de Mayo y Francia (1650) San Martín, Argentina}
\affiliation{Departamento de Energía Solar, Centro Atómico Constituyentes (CNEA), Av. Gral Paz 1499, San Martín (Buenos Aires), Argentina}

\author{Sebastian Bustingorry}
\affiliation{Instituto de Nanociencia y Nanotecnología CNEA-CONICET, Centro Atómico Bariloche, Av. E. Bustillo 9500, (R8402AGP) S. C. de Bariloche,
Río Negro, Argentina}

\author{Javier Curiale}
\affiliation{Instituto de Nanociencia y Nanotecnología CNEA-CONICET, Centro Atómico Bariloche, Av. E. Bustillo 9500, (R8402AGP) S. C. de Bariloche,
Río Negro, Argentina}
\affiliation{Instituto Balseiro, Universidad Nacional de Cuyo - CNEA, Av. Bustillo 9500, R8402AGP, S. C. de Bariloche, Rio Negro, Argentina}

\author{Mara Granada}
\email{granadam@cab.cnea.gov.ar}
\affiliation{Instituto de Nanociencia y Nanotecnología CNEA-CONICET, Centro Atómico Bariloche, Av. E. Bustillo 9500, (R8402AGP) S. C. de Bariloche, Río Negro, Argentina}

\date{\today}

\begin{abstract} 
Understanding the effect of fabrication conditions on domain wall motion in thin films with perpendicular magnetization is a mandatory issue in order to tune their properties aiming to design spintronics devices based on such phenomenon. In this context, the present work intends to show how different growth conditions may affect domain wall motion in the prototypical system Pt/Co/Pt. The trilayers were deposited by dc sputtering, and the parameters varied in this study were the Co thickness, the substrate roughness and the base pressure in the deposition chamber.
Magneto-optical Kerr effect-based magnetometry and microscopy combined with X-ray reflectometry, atomic force microscopy and transmission electron microscopy were adopted as experimental techniques. 
This permitted us to elucidate the impact on the hysteresis loops and on the domain wall dynamics, produced by different growth conditions. 
As other authors, we found that Co thickness is strongly determinant for both the coercive field and the domain wall velocity. On the contrary, the topographic roughness of the substrate and the base pressure of the deposition chamber evidence a selective impact on the nucleation of magnetic domains and on domain wall propagation, respectively, providing a tool to tune these properties.
  
\end{abstract}

\maketitle

\section{Introduction}

Nowadays, magnetization reversal and magnetic domain wall (DW) dynamics are key topics in the field of electronics and spintronics.
Even if the discovery of current-driven DW motion by spin transfer torque paved the way for novel DW based devices, there are still many hurdles to overcome before this technology becomes massively used.
The control of DW short and long-term stability and displacement is critical for potential applications such as magnetic storage and emerging DW-based spintronic devices \cite{HirohataReviewSpintronicJMMM2020, LuoNat2020, JourPhysD_Roadmap2020}. 
Moreover, since the field- and current-driven DW dynamics share many similarities \cite{PRBUniversalDWMotion}, all the knowledge achieved from either side contributes to the full understanding of the physics of the DW dynamics. 
In this context, the most promising systems are those presenting a dominant perpendicular magnetic anisotropy (PMA)\cite{HirohataReviewSpintronicJMMM2020, broeder_magnetic_1991}. 
Systems having a dominant PMA with out-of-plane magnetization offer better scale-down capacity and require lower current to induce magnetization switching, among other advantages for spintronics applications\citep{dieny_perpendicular_2017}.

Pt/Co/Pt is a prototypical system in which PMA was early observed and the magnetic DW motion discussed \citep{lemerle_domain_1998}. Since then, a large number of works have reported on how to 
tune its properties by controlling different deposition parameters as the buffer layers, thickness of the Co and the Pt layers and also performing different annealing treatments
\citep{ferre_JMMM_1999, metaxas_creep_2007, kanak_influence_2007, emori_optimization_2011, chowdhury_effect_2012, je_PRB_2013, lee_effects_2013, lim_effect_2015, jue_nature-mat_2015, bersweiler_impact_2016, wang_2020}.
Co-based simple stacks continue to be at the focus of intense research activity since the renewed interest on the Dzyaloshinskii-Moriya interaction \citep{fert_skyrmions_2017}. In this regard, recent studies of domain expansion under in-plane field in sputtered Pt/Co/Pt, have demonstrated the relevance of modifying the interfaces by controlling the deposition conditions. The Ar pressure during the deposition of the top Pt layer \citep{Lavrijsen_2015}, the deposition pressure and the substrate temperature \citep{wells_2017} were shown to have an impact on the strength of the PMA and on the asymmetric velocity of magnetic bubble expansion under in-plane applied magnetic field. 
The above mentioned studies have proved that DW velocity is very  sensitive to the deposition parameters. In fact, the precise deposition conditions may not be reproducible in different deposition systems \citep{Lavrijsen_2015}, so that reporting on general trends turns to be more useful than paying attention to the precise parameters (base and deposition pressures, substrate temperature, substrate-to-target distance, layers thicknesses, etc.), when comparing samples from different groups.

Within this framework, in this study we contribute to clarifying the role of some deposition conditions on the properties of Pt/Co/Pt films, mainly through the domain wall propagation analysis. 
We assess the impact of the substrate roughness and base pressure, compared to the one produced by the Co thickness, on the DW propagation velocity and coercive field. Whilst substrate roughness mainly affects the nucleation of magnetic domains, base pressure in the deposition chamber predominantly impacts on DW propagation.

\section{Samples and Methods}

Pt/Co/Pt films were deposited by dc magnetron sputtering at room temperature in a  $(2.8 \pm 0.1) \times$10$^{-3}$ Torr Ar atmosphere. 
Pt and Co targets were sputtered with 20 W and 10 W, respectively. The deposition rates were (1.25 $\pm$ 0.05) $\text{\AA}$/s for platinum and (0.38 $\pm$ 0.08) $\text{\AA}$/s for cobalt, for a distance from substrate to target of 86 mm. The trilayers studied in this work typically consist of Pt(8 nm)/Co($d_\text{Co}$)/Pt(4 nm) with $d_\text{Co}$ ranging between 0.4 and 0.7 nm. Four diffent substrates were used in this work. Two of them are (001) oriented Si wafers from different manufacturers and different production years 
: MTI\textregistered , 1998 (S1) and Crystal\textregistered , 2012 (S2).
Although they are nominally the same material, they differ in their surface topography.
We also used (001) oriented SrTiO$_3$ (STO) and thermally oxidized S1 with a 100 nm thick SiO$_2$ layer (SiO$_2$).

Systematic studies of the structural properties, coercive field and domain wall propagation were performed on three batches of samples:\vspace{2mm}

\noindent \textbf{Series A: different Co thicknesses.} Pt/Co($d_{\text{Co}}$)/Pt samples with varying Co deposition times ($t_{\text{Co}}$ = 12, 15 and 18 s), which resulted in different Co thicknesses ($d_\text{Co}$). This series was deposited on silicon (S1) substrates. The base pressure ranged between 9.0 and 9.3 $\times 10^{-7}$ Torr.\vspace{2mm}

\noindent \textbf{Series B: different substrates.} Pt/Co/Pt trilayers deposited on three different substrates: silicon (S2), SrTiO$_3$ (STO) and SiO$_2$. The deposition time for Co was $t_\text{Co}$ = 18 s and the base pressure ranged between 6.7 and 7.3 $\times 10^{-7}$ Torr.\vspace{2mm}

\noindent \textbf{Series C: different base pressures.} Pt/Co/Pt trilayers deposited on Si (S1) substrates with $t_\text{Co}$ = 15 s using different base pressures in the sputtering chamber, 3$\times$10$^{-6}$ and 1$\times$10$^{-5}$ Torr, before the deposition procedure. Since the deposition takes place with the same Ar pressure as the other series, we assume that the initial base pressure may affect the cleanness of the environment rather than the growth dynamics.\vspace{2mm}

It is worth to notice that Series B does not include a trilayer deposited on S1 substrate. The unexpected identification of the base pressure as a relevant parameter, plus a careful analysis of the reproducibility of DW velocity curves (see the Appendix), prompted us to revise the definition of the batches of samples. After having established highly restrictive criteria for the present work, the sample deposited on S1 with the same thickness as those of Series B, can not be included in that Series.

Transmission electron microscopy (TEM) images were acquired with a Philips CM200 microscope using an acceleration voltage of 200 keV. Samples for TEM observations were prepared by focused ion beam (FIB) with a FEI Helios Nanolab 650 dual beam system. A lamella was extracted from the bulk sample by ion milling and then transported to a copper grid with micromanipulators, where it was fixed at its final position with a local Pt deposition. The lamella was then thinned to sub-100 nm thickness and a final cleaning process was performed with 5 kV ion beam acceleration voltage. 

Atomic force microscopy (AFM) images were acquired in tapping mode with a Veeco Dimension 3100 NanosScope.

X-ray reflectometry (XRR) experiments were performed with Cu-K$\alpha$ radiation ($\lambda = 1.54$ \AA) using an Empyrean PANalytical System in Bragg-Brentano configuration. 

Magneto-optical Kerr effect (MOKE) magnetometry in the polar configuration was used to measure out-of-plane magnetization loops. The experimental curves presented in this work were acquired with a magnetic field sweeping rate of 220 Oe/s.

The magnetic-field-driven domain wall (DW) motion was studied in a home-made MOKE microscope in the polar configuration (PMOKE). The most relevant features of the used microscope are the \textit{Olympus LMPLFLN} series objectives (20$\times$ and 5$\times$), a high-brightness red LED with a dominant wavelength 637 nm, two Glan-Thompson polarizers, a 14 bit CCD from \textit{QImaging Corp.} and the illumination set in a K\"{o}hler configuration.
PMOKE microscopy images consist of regions with two different gray levels, that correspond to magnetic domains with the magnetization pointing in opposite directions perpendicular to the sample plane.
Due to the weak magnetic contrast in a typical PMOKE image of the studied samples, it is better to work with differential images. Either an image of the saturated sample is subtracted as a background, or consecutive micrographs are subtracted from each other.
The acquisition and analysis procedures followed in this work are described in detail in Ref. [\onlinecite{APL2018_Quinteros}]. The sample is first saturated and then a magnetic field pulse with the opposite polarity is applied to produce domain nucleation. In order to study the DW dynamics, a series of square magnetic field pulses of intensity $H$ and duration $\Delta t$ are applied and PMOKE images are acquired after each pulse. The DW velocity is computed as $v = \Delta x / \Delta t$, where $\Delta x$ is the distance traveled by the DW between consecutive images. By reproducing this procedure for different magnetic field intensities, we obtain $v(H)$ curves. 

All the results presented in this work were obtained at room temperature.

\section{Results}

\subsection{Structural characterization}

\begin{figure}[h!]
    \centering
    \includegraphics[width=0.8\columnwidth]{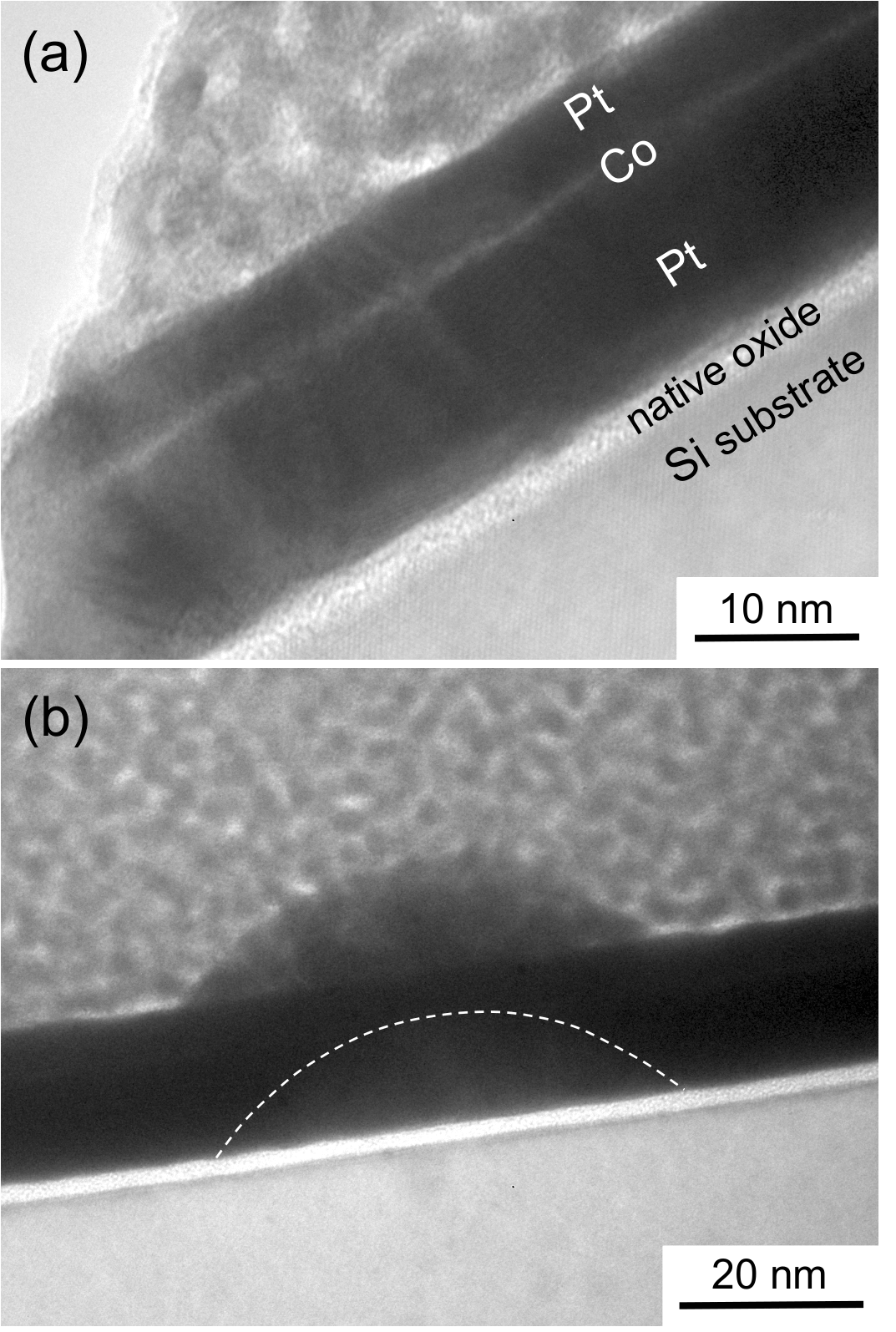}
    \caption{Transmission electron microscopy images of a lamella of a Pt/Co/Pt film on a Si S1 substrate prepared with focused ion beam. (a) Continuity of the Co layer is demonstrated. (b) A bump is observed in both the substrate and the trilayer. The dashed line is a guide to the eye, delimiting the brighter region identified as a bump on the substrate.}
    \label{fig:TEM_image}
\end{figure}

Figure \ref{fig:TEM_image} shows TEM images of a Pt/Co/Pt lamella. Since the contrast is related to the atomic mass, Co can be distinguished in Fig. \ref{fig:TEM_image}(a) as a brighter line between the two dark Pt layers. The Si substrate and its native oxide layer are indicated. On top of the trilayer, granular Pt deposited during the FIB processing can be recognized. Continuity of the Co film was observed along the whole extension of the lamella. This observation allows us to rule out all possible effects coming from discontinuities of the Co layer \citep{charilaou_magnetic_2016}.
Complementarily, in Fig. \ref{fig:TEM_image}(b) a bump is observed in both the film and the substrate, which indicates that the trilayer copies the topographical defects of the substrate. The dark portion of the image corresponds to the Pt/Co/Pt stack, the Co layer being hardly visible in this magnification. The bump on the substrate is distinguished as a brighter region within the dark gray of the Pt/Co/Pt film, indicated with a dashed line. 

\begin{figure*}[t!]
    \centering
    \includegraphics[width=0.9\textwidth]{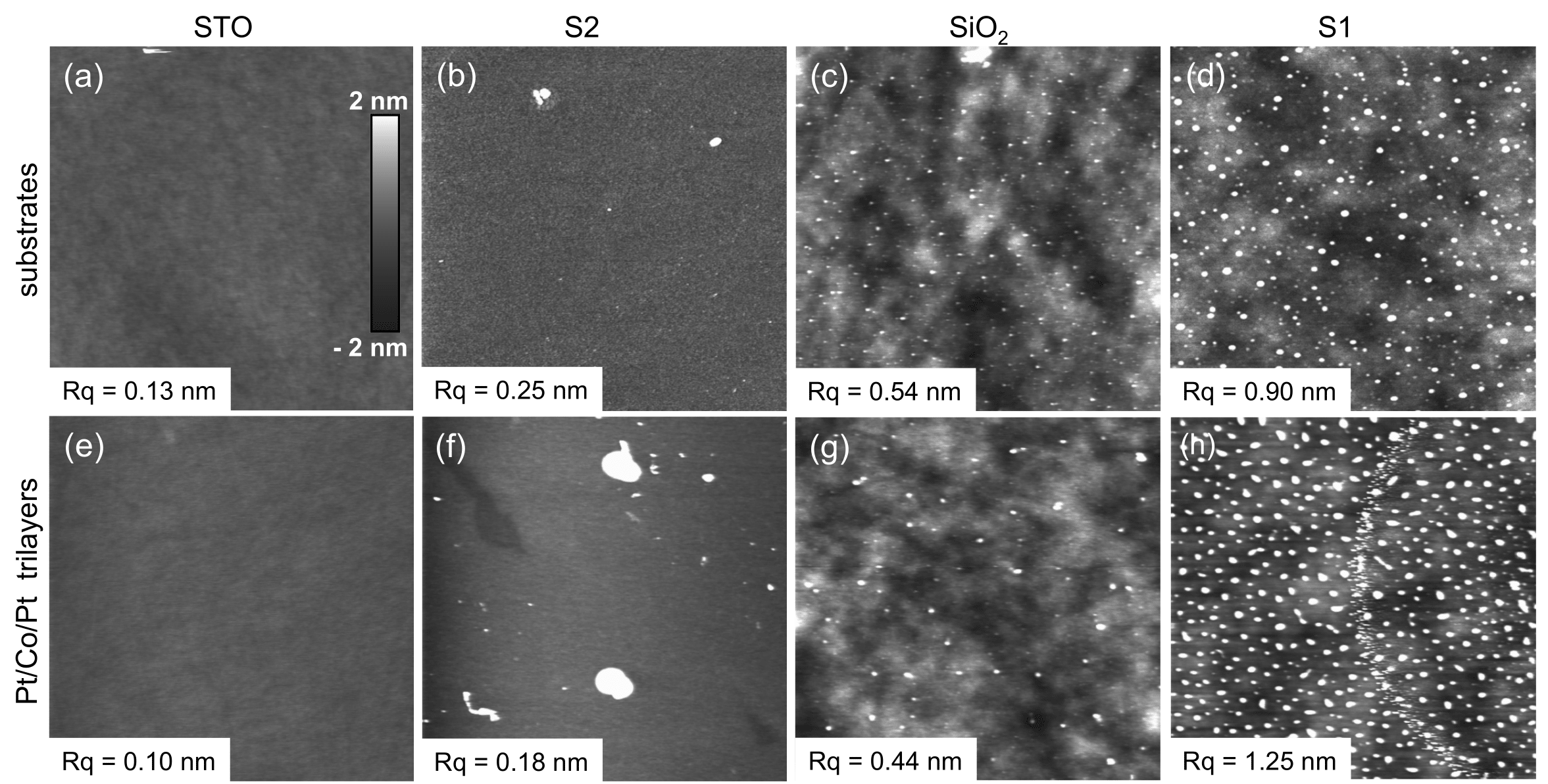}
    \caption{Atomic force microscopy images of the surface topography. Above: the substrates STO (a), S2 (b), SiO$_2$(c) and S1 (d). Below, (e)-(h): Pt/Co/Pt trilayers selected from Series A and B, with the same Co thickness ($d_\text{Co}$ = 0.64) nm, deposited over the different substrates. All the images have 5 $\mu$m $\times$ 5 $\mu$m area and are displayed in the same gray scale, with 4 nm amplitude. The roughness $Rq$ was computed using a (2 $\mu$m)$^2$ selected area from each sample. In the case of the trilayer on S2 (f), the big defects were intentionally avoided for the computation of $Rq$. (b) and (d) are reprinted from [\onlinecite{APL2018_Quinteros}], with the permission of AIP Publishing.}
    \label{fig:AFM}
\end{figure*}

AFM images are presented in Fig. \ref{fig:AFM} for the different substrates [panels (a)-(d)] and Pt/Co/Pt trilayers deposited on each of them [panels (e)-(h), respectively]. For different substrates, different density of defects and roughnesses were observed. To quantify the roughness, $Rq$ (the RMS value of the height deviation from the average plane) was computed for (2 $\mu$m)$^2$ selected areas. The images in Fig. \ref{fig:AFM} are ordered by increasing roughness from left to right.
The SiO$_2$ substrate, obtained by thermally oxidizing a S1 wafer, presents a reduced roughness after the process. Figure \ref{fig:AFM} shows that the Pt/Co/Pt trilayers reproduce the topography of the substrates beneath, as suggested also by TEM observations.

\begin{figure*}[ht!]
    \centering
    \includegraphics[width=0.9\textwidth]{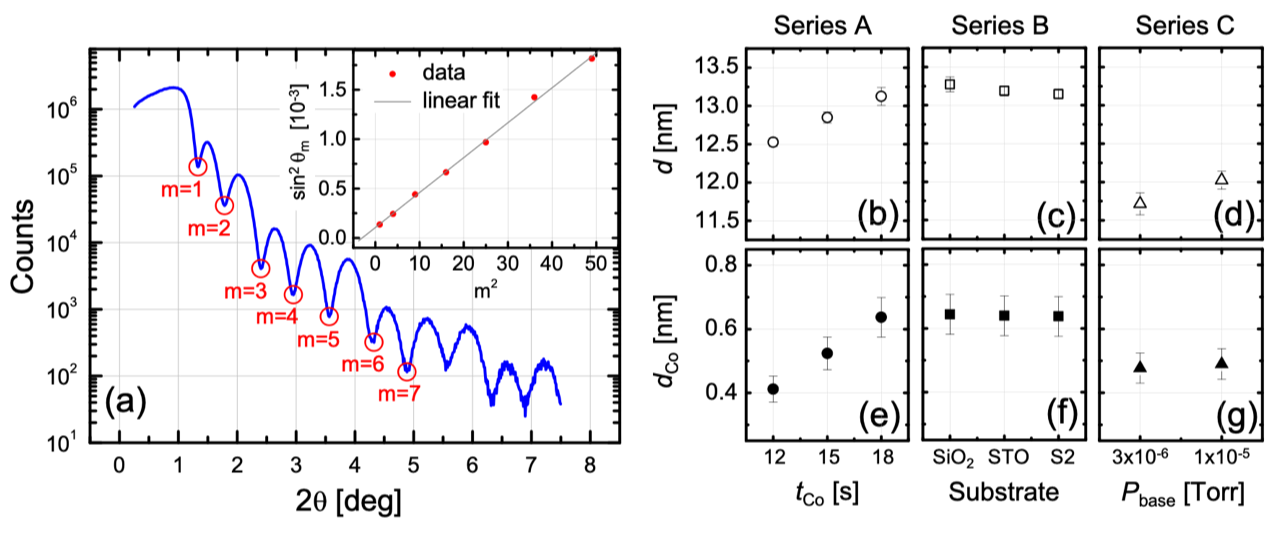}
    \caption{(a) X-ray reflectometry of a Pt/Co/Pt sample with $t_\text{Co}$ = 18 s deposited on S1 substrate (Series B). The angular positions of the minima are indexed. The inset shows $\sin ^2 \theta_m$ vs. $m^2$ (being $\theta_m$ the angles for XRR minima and $m$ the index of each minimum). The line is a linear fit to the experimental data; the slope is used to compute the total thickness of the samples by means of Eq. (\ref{eq.Bragg}). The total thicknesses computed for samples with (b) different Co thicknesses, (c) different substrates and (d) different base pressures are presented. The respective Co thicknesses, displayed in (e)-(g), were estimated using Eq. (\ref{eq.dCo}) as described in the text.}
    \label{fig:XRR}
\end{figure*}

Figure \ref{fig:XRR}(a) shows a measured XRR curve for a trilayer with total thickness of 13.13 nm deposited on S1 substrate. The periodicity of the oscillations (Kiessig fringes) in the XRR curves is determined by the total thickness of the sample \cite{HolyXR}. When the electronic density of the film is higher than that of the substrate, as in the present case, the total thickness $d$ of the film is related to the angular positions of the minima $\theta_m$ through the modified Bragg law \cite{Nakamura_1992}

\begin{equation}
\sin ^2 \theta_m = 2 \delta + m^2 (\lambda/2d)^2 \text{,}
 \label{eq.Bragg}
\end{equation}

\noindent where $1- \delta$ is the real part of the refractive index of the film and $m$ the order of each minimum, indicated in Fig. \ref{fig:XRR}(a). The inset of Fig. \ref{fig:XRR}(a) shows the relationship between $\sin^2 \theta_m$ and $m^2$; the total thickness $d$ of the sample is obtained from the slope of the linear fit to the experimental data, by using Eq. (\ref{eq.Bragg}). This procedure was employed to measure the total thickness of all the studied samples, the results are displayed in Fig. \ref{fig:XRR}(b)-(d). Our deposition method gives a dispersion in the measured total thickness of up to 10 \%, when comparing samples with the same deposition times but from different batches (see the Appendix for further discussion about reproducibility).
Since the Co thickness ($d_\text{Co}$) is one of the parameters whose influence on the magnetic properties we intend to study, we carefully estimated this quantity for each trilayer. In order to compute $d_\text{Co}$, we used the previously calibrated deposition rates $r_\text{Pt} = $ (1.25 $\pm$ 0.05) $\text{\AA}$/s for Pt and $r_\text{Co} = $ (0.38 $\pm$ 0.08) $\text{\AA}$/s for Co, and we assumed that the ratio $R$ between those rates remains unchanged with a value  $R = r_\text{Pt} / r_\text{Co}$ = 3.3 $\pm$ 0.4. Under that assumption, knowing the deposition times $t_\text{Pt}$ and $t_\text{Co}$ for each single layer and using the experimental total thicknesses $d$ presented in Fig. \ref{fig:XRR}(b)-(d), we estimated the Co thickness $d_\text{Co}$ for each sample as

\begin{equation}
d_\text{Co} = \frac{t_\text{Co} ~d}{t_\text{Co} + R ~t_\text{Pt}} \text{ .}
 \label{eq.dCo}
 \end{equation}

\noindent The results, presented in Fig. \ref{fig:XRR}(e)-(g), confirm that while Series A has varying $d_\text{Co}$, Series B and Series C trilayers have the same Co thickness within each series.
Thus, we can state that any effects due to varying Co thickness will only be observed within Series A.

\subsection{Domain wall velocity and coercive field}

\begin{figure*}[htbp]
    \centering
    \includegraphics[width=0.8\textwidth]{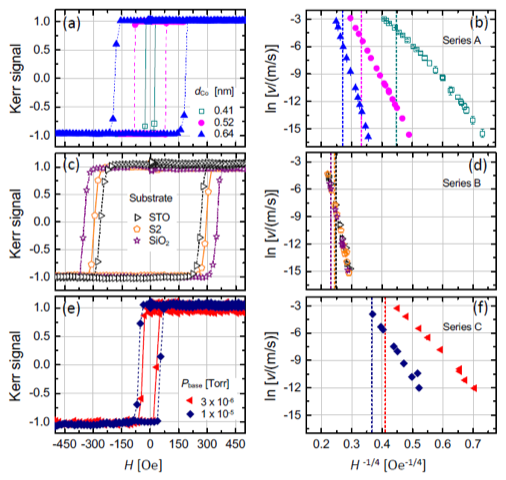}
    \caption{Normalized magnetization loops obtained by Kerr magnetometry (left panels) and DW velocity curves displayed as $\ln v$ vs. $H^{-1/4}$ (right panels) grouped by series of samples. Different symbols and colors are used to identify the results corresponding to each sample. The vertical dashed lines in the right panels indicate the coercive field obtained from the magnetization loop of each sample (left panels). All the measurements were carried out at room temperature and in the polar configuration, i. e., with the magnetic field applied perpendicular to the sample plane. (a) Magnetization loops and (b) DW velocity for \textbf{Series A}: samples with different $d_\text{Co}$ deposited over S1 substrates with $P_\text{base} = (9.1 \pm 0.2) \times 10^{-7}$ Torr. (c) Magnetization loops and (d) DW velocity for \textbf{Series B}: Pt/Co/Pt films deposited on different substrates, with $d_\text{Co} = (0.64 \pm 0.06)$ nm and base pressure before deposition $P_\text{base} = (7.1 \pm 0.2) \times 10^{-7}$ Torr. (e) Magnetization loops and (f) DW velocity curves for \textbf{Series C}: samples desposited with different $P_\text{base}$ on S1 substrates with cobalt thickness $d_\text{Co} = (0.48 \pm 0.05)$ nm.  (a) and (b) are reprinted from [\onlinecite{APL2018_Quinteros}], with the permission of AIP Publishing.}
    \label{fig:Mag}
\end{figure*}

In the following, we present a study of the variations in the DW velocity and coercive field produced by changing either the Co thickness (Series A), the substrate (Series B) or the base pressure in the deposition chamber (Series C).
The DW velocity and magnetization curves obtained for all the samples are displayed in Fig. \ref{fig:Mag}.
The out-of-plane magnetization loops have a square shape, due to predominant PMA in all the samples. The DW velocity curves are presented as $\ln(v)$ vs. $H^{-1/4}$ plots. The fact that these curves present a linear behavior is compatible with the DW movement taking place within the creep regime\cite{nattermann_scaling_1990,lemerle_domain_1998,chauve_creep_2000,jeudy_universal_2016}.
We have evaluated the degree of reproducibility of our samples, and we confirmed that the variations in DW velocity vs. magnetic field in Fig. \ref{fig:Mag} are larger than the dispersion due to reproducibility limitations (see the Appendix for more details). The vertical dashed lines in the right panels of Fig. \ref{fig:Mag} indicate the coercive field for each sample obtained from the magnetization loops in the left panels.
The DW velocity at the coercive field is in the range $10^{-3}\,\mathrm{m/s} < v < 1\,\mathrm{m/s}$. Therefore, taking into account typical times associated to the magnetization measurements, once the domains are nucleated at $H_C$ they rapidly grow, giving rise to the square shape of the magnetization loop. This suggests that the coercive field $H_C$ is mainly associated to the process of nucleation of magnetic domains, while the measured DW velocity is due to the domain growth process.

The magnetization loops for Series A [Fig. \ref{fig:Mag}(a)] show that the coercive field increases with increasing thickness. This in agreement with the results of Metaxas \textit{et al.}\citep{metaxas_creep_2007} on Pt/Co/Pt, but opposite to what Su \textit{et al.} \citep{su_evolution_2016} obtained for Co/Ni stacks, which suggests that this behavior is material-dependent. In addition, Fig. \ref{fig:Mag}(b) demonstrates that DW velocity changes dramatically with the Co thickness: the DW velocity at 15 Oe is 2.5 10$^{-3}$ m/s for the $d_\text{Co}$ = 0.41 nm sample against 1.1 10$^{-7}$ m/s for the $d_\text{Co}$ = 0.52 nm sample. 
It is clear that Co thickness is strongly determinant for both the coercive field and the DW velocity. On the contrary, the parameters modified in the other two series evidence a selective impact on the magnetization reversal and DW propagation.

The magnetization loops presented in Fig. \ref{fig:Mag}(c) evidence a variation of the coercive field in samples of Series B. It is shown that the coercive field increases with increasing roughness, as also observed in similar samples by other authors\citep{kanak_influence_2007}.
Figure \ref{fig:Mag}(d) shows that the field-driven DW propagation in the thermally activated creep regime, is the same for all these samples. 
This indicates that the topographical defects contributing to the $Rq$ are not acting as relevant pinning centers for the DW motion. 
This could be associated to different characteristic length-scales of the defects and DW.
The typical sizes of the defects observed by AFM are roughly 10 nm high with tens to hundreds of nanometers in-plane diameter, separated by a few hundreds of nm. 
On the other hand, the width of the DW for a similar system was estimated by Gorchon \emph{et al.} \citep{gorchon_pinning-dependent_2014} as $\Delta \sim$ 5-10 nm. Then, it is not surprising that the DW velocity is not affected by the topography, since the surface presents smooth variations in the scale of the DW width.



Finally, two different base pressures were used (Series C), both of them orders of magnitude lower than the Ar pressure required for plasma ignition. 
In this case, the magnetization loops are presented in Fig. \ref{fig:Mag}(e) and the DW velocity is shown in Fig. \ref{fig:Mag}(f). 
We can note that the coercive field is not very sensitive to the base pressure while, on the contrary, the effect of this parameter on the DW dynamics is much more significant. Fig. \ref{fig:Mag}(f) demonstrates that the DW motion is strongly affected by the initial condition of the deposition process: the DW velocity is higher in the case of a cleaner environment before the introduction of Ar. If we choose a fixed field value (e.g. 16 Oe), velocities that differ in more than two orders of magnitude are found (4.1 10$^{-3}$ m/s for 3 10$^{-6}$ Torr and 2.8 10$^{-5}$ m/s for 1 10$^{-5}$ Torr). 
This indicates that the DW velocity in the creep regime is not only influenced by the Ar pressure during deposition, as previously pointed out by other authors \citep{emori_optimization_2011,park_effect_2016,Lavrijsen_2015}, but also by the base pressure prior to deposition. This parameter of the growth process defines the degree of purity of the gas inside the chamber which impacts on the quality of the obtained materials. This would affect the propagation of the DW by modifying the density and/or strength of pinning centers.


\section{Discussion and Conclusions}

To summarize, in this work we studied the impact of deposition conditions on the coercive field and DW velocity in perpendicularly magnetized Pt/Co/Pt stacks grown by dc sputtering. We deposited three different batches of samples varying different deposition parameters: the Co thickness (Series A), the substrates roughness (Series B) and the base pressure in the deposition chamber (Series C).
A common feature to all the samples is that the Pt/Co/Pt trilayers reproduce the topography of the substrates.
We have observed that changes on each of the studied parameters affect the coercive field and the DW velocity in different ways. In this regard, from the results obtained for Series A we can state that the Co thickness affects both the coercive field and the DW velocity significantly.
Our results corresponding to Series B, suggest that the main effect of changing the substrate is observed on the coercive field, thus a change in the topographical roughness of the samples impacts on the nucleation process. 
At the same time, the DW propagation is roughly insensitive to the surface topography. The fact that the topographical defects do not contribute to the pinning of the DW could be associated to different characteristic length-scales of both objects. Indeed, as the observed topographical defects have typically 10 nm height and more than 50 nm in diameter, the surface presents smooth variations in the scale of the DW width ($\Delta \sim$ 5-10 nm).
The opposite is observed in Series C: changing the base pressure in the deposition chamber strongly affects the DW propagation process but not so much the coercive field. For example, oxygen content in the chamber, proportional to the base pressure, might result in the oxidation of cobalt, thus locally changing the magnetic properties and originating pinning centers for DW motion.
Then, the quality of the thin magnetic film at the microscopic level would modify the density and/or strength of pinning centers but would not affect appreciably the nucleation energy.

The main contribution of this work is to evidence that the nucleation of magnetic domains and the propagation of DW, which rely on different physical mechanisms, can be modified independently by changing different deposition parameters. This allows to tune the coercive field and DW velocity of Pt/Co/Pt samples, by choosing an appropriate combination of deposition parameters.
Noteworthy, parameters such as the base pressure in the deposition chamber, which is not usually reported, might have a significant impact on DW dynamics.
Then, this work also contributes to understand the dispersion of magnetic properties, in particular DW velocity, of nominally equivalent samples. 

\section*{Acknowledgments}

We acknowledge grants from Agencia Nacional de Promoción Científica y Tecnológica (PICT 2014-2237, PICT 2016-0069, PICT 2017-0906) and Universidad Nacional de Cuyo (UNCuyo 06/C561). 
CQ also acknowledges financial support from NWO’s TOP-PUNT grant 718.016002.

\section*{Appendix}

\begin{figure}[h!]
    \centering
    \includegraphics[width=0.9\columnwidth]{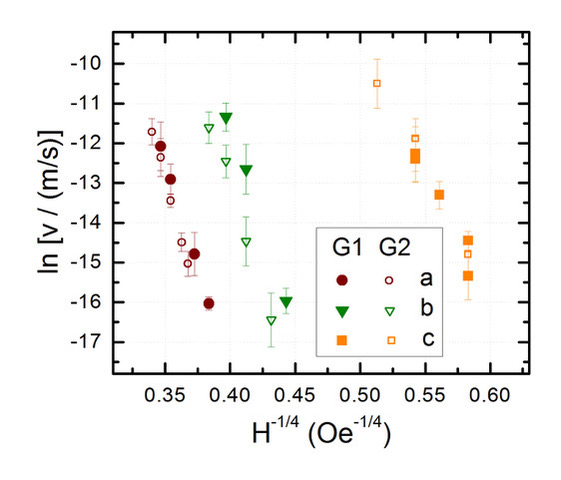}
    \caption{DW velocity curves for two groups of samples grown under the same conditions (G1 and G2), each of them composed of three Pt/Co/Pt samples, with different Pt and Co thicknesses (a, b and c).}
    \label{fig:reproducibility}
\end{figure}

We assessed the degree of reproducibility of the DW velocity in the creep regime for two reasons. On the one hand, DW dynamics is our main interest, and on the other hand, the DW velocity is a property that is strongly affected by subtle changes in the sample.
In Fig. \ref{fig:reproducibility}, we present the DW velocity measured for six Pt/Co/Pt trilayers deposited on the same substrate.
We first deposited batch G1, consisting of three samples: a, b and c with different Pt and Co thicknesses. We then repeated exactly the same deposition procedure and obtained the second batch, G2. As demonstrated in Fig. \ref{fig:reproducibility}, samples grown under the same conditions (including base pressure and substrate roughness) have equivalent DW velocity. 

\bibliography{biblio}

\end{document}